\documentstyle[12pt,epsf,tighten,prd,aps]{revtex}
\begin{document}
\begin{center}
{\bf Exact inflationary solutions}\\
\vspace*{10pt}

{\bf Gabriella Piccinelli}\footnote{e-mail: gabriela@astroscu.unam.mx}\\
Instituto de Astronom\'{\i}a\\ Universidad Nacional Aut\'onoma de M\'exico
\\Apartado Postal 70--264, 04510, M\'exico D. F., M\'exico.\\
{\bf Tonatiuh Matos}\footnote{Permanent Address: Departamento de
F\'{\i}sica, CINVESTAV--IPN, 
Apartado Postal 14-740, 07000, M\'exico D.F., M\'exico.\\
 e-mail: tmatos@fis.cinvestav.mx},\\ 
Instituto de F\'{\i}sica y Matem\'aticas\\
Universidad Michoacana de San Nicol\'as de Hidalgo\\
Apartado Postal 2-82, Morelia 58040, Michoac\'an, M\'exico.\\and\\
{\bf Merced Montesinos}\footnote{Current Address: Department of 
Physics and Astronomy,
University of Pittsburgh, Pittsburgh, PA 15260, USA.\\
e-mail: merced@fis.cinvestav.mx}\\
Departamento de F\'{\i}sica\\ Centro de Investigaci\'on y de Estudios 
Avanzados del IPN\\ Apartado Postal 14-740, 07000, M\'exico D.F.,
M\'exico.
\end{center}
\begin{abstract}
We present a new class of exact inflationary solutions for the 
evolution of a universe with spatial curvature, filled with a perfect
fluid, a scalar field with potential
$V_{\pm}(\phi)=\lambda(\phi^2\pm\delta^2)^2$ and a cosmological constant
$\Lambda$. With the $V_+(\phi)$ potential and a negative cosmological
constant, the scale factor experiments a graceful exit. 

We give a brief discussion about the physical
meaning of the solutions. 
\end{abstract}
\baselineskip 20pt

\section{INTRODUCTION}

Inflation is a useful concept for cosmology, both as a theory of initial 
conditions and providing the physical conditions for seeding the universe 
large scale structure (see e.g. Linde \cite{linde}). 
After the old--inflationary model \cite{Guth}, where inflation ended through 
a first--order phase transition in which the true--vacuum bubbles never 
percolated, inflation has been traditionally achieved for a slow 
rollover scalar potential $V(\phi)$, assuming it is the dominant term in the 
total energy density and neglecting the contributions of the kinetic energy 
${1\over 2} \dot\phi^2$, in order to have a de Sitter-like inflation, with a 
(quasi) exponential solution. Subsequently, recognizing that an 
inflationary expansion can be defined just by $\ddot a > 0$ (where $a$ is 
the universe scale factor), models with a steeper potential have been 
considered, leading to power--law inflation (see e.g. \cite{lucchinm}). These 
models can be related, by a conformal transformation, to a 
scalar--tensor gravity theory with a standard inflationary potential, 
thus introducing modifications in the gravitational sector too \cite{alguien}.

Many inflationary models have been proposed up to now 
(for a review see e.g. \cite{olive}), all of them with a more or less 
pronounced fine--tuning problem. 
Recently, in order to keep pace with astrophysical observations that 
seem to point to low--$\Omega$ values ($\Omega$ is the universe density 
parameter), a double--round inflation model that leads to $\Omega < 1$ 
has been extensively developed (see \cite{artlinde} and references there in).

As we mentioned, in the usual approach to the early universe evolution, only 
the  conjectured dominating energy density during each epoch is kept as the 
source for the universe expansion, neglecting all the sub--dominant 
components. In this way, the inflationary solutions 
use to emerge from many assumptions that considerably simplify the 
equations. A work from Chimento and Jakubi \cite{chim} 
analyses some exact solutions in a many--components
cosmology, showing the limitations imposed by the slow--rollover 
approximation. They find that the fluid source and/or the spatial 
curvature may dominate the early universe evolution and inhibit 
inflation. Schunck and Mielke \cite{mielke,milke1} present a method for 
finding exact solutions using the Hubble parameter as a time 
coordinate; they find some exact inflationary solutions and compare them 
with COBE results in some approximation.

Another possibility for the cosmological scenario is the existence of a 
cosmological constant term. It has been widely exploited in structure 
formation models and it has been mainly considered in order to extend the 
Universe lifetime \cite{carroll}. This term is expected to dominate at later 
epochs but it is anyhow a possible component of the very early Universe.

An important issue in constructing a successful inflationary 
model is to find a fundamental theory that naturally contains it. 
An appealing framework for inflationary fields is higher--dimensional 
theories. There, like in Kaluza-Klein (KK) and in superstring 
theories (SS), the scalar field --the dilaton-- is a natural component of 
the field equations.
Nevertheless, inflationary models from higher--dimensional theories have 
encountered serious problems : either they do not inflate enough (see 
\cite{picci} and references therein) or they have a graceful 
exit problem, as in \cite{Levin}. In \cite{broso} the fermionic part, 
together with a Higgs potential, have been added to an eight-dimensional 
theory in order to recover the standard model from a KK theory. Thus, 
KK becomes an n--dimensional model for the standard model, coupled to
gravity. If we introduce the FRW-metric in a higher-dimensional KK theory,
the fermionic sector vanishes (gauge fields define a privileged direction) 
and the scalar sector reduces to one scalar field \cite{cho}.

On another hand, a general feature of superstring cosmology is that until 
the dilaton 
field settles near a minimum of its potential, its kinetic energy 
dominates the potential energy, preventing inflation. In this context, 
the best possibility 
should be that the dilaton settles down and inflation is driven by other 
fields.  Nevertheless, for $\phi$ to be trapped in a minimum of its 
potential, some very strict constraints on the initial conditions have 
to be satisfied. Moreover, the dilaton potential typically has minima 
with negative vacuum energy density and this carries many problems for 
cosmology: the difficulties to avoid a minimum with negative cosmological 
constant and an unacceptably high variation of the gravitational coupling 
(see e.g. \cite{Brustein}). 

Successful inflation may occur due to chiral fields, under certain 
conditions, once the dilaton and moduli fields are stabilized by demanding an 
S--duality \cite{Font} invariant potential \cite{axel}. Another viable model, 
called false vacuum inflation \cite{Cop}, have been built in a class of 
supergravity models that follow from orbifold compactification of 
superstrings, albeit again with significantly constrained parameters. It 
requires two scalar fields --one rolling and the other trapped in a false 
vacuum state-- and the false vacuum energy can give the dilaton a suitable 
minumum during inflation. 

Another interesting possibility arises when the 
duality symmetry is invoked. Gasperini and Veneziano \cite{venez} use the 
scale factor duality (SFD) and the time inversion of the string effective
equations to deduce the existence of a scenario named ``pre-big-bang"
which is associated, by SFD and time inversion, to the ``post-big-bang"
epoch, namely, the actual epoch. 
In this context, the big bang epoch means the transition between these two 
dual phases. In \cite{luca}, starting from the multidimensional Brans-Dicke 
theory with a perfect fluid, solutions to the equations of motion were found 
which contain a singularity in the curvature.
This approach also admits solutions
without singularities when a self--dual potential is introduced for the 
dilaton field. 
The scale factor grows non-monotonically  
through three phases of accelerated expansion, contraction, expansion, before 
the final decelerated expansion (standard cosmological model). Inflation can 
be accomodated in these three phases of accelerated evolution.

Keeping in mind all these possibilities for the universe content and 
features --dimensions, curvature, different inflationary fields and 
episodes--, in this work we kept all the terms possibly envolved in the 
early universe dynamics and
we were able to find exact inflationary solutions for a closed universe and 
different signs of the cosmological constant. 
In each case, $\Lambda$ and $\rho_0$ will be defined 
by the parameters of the theory. The requirement that $\rho > 0$ will 
put constraints on the allowed values of the 
parameters and hence on the form of the solutions. 

In section II we briefly present the model and in the two following sections 
we give two classes of inflationary solutions with a brief discussion on 
their main features and a comparison to previous work which can be found in 
the literature and was mentioned in this introduction. Section V contains a 
summary of this work.
\section{THE MODEL}

In the KK theory the compactification of the extra 
dimensions yields a classical unified theory of gravitation and the 
Yang-Mills fields. The dilaton is defined during the process of 
dimensional reduction, as a function of the scale factor of internal 
dimensions.
The simplest KK model consists on a five-dimensional space projected 
into four-dimensions. Then, the unified $U(1)$ Yang-Mills field is 
associated to electromagnetism and the dilaton may play the role of the 
scalar field required for inflationary models. 

After compactification, Kaluza--Klein and superstring theories are both 
special cases of a more general Lagrangian given by 
\begin{eqnarray}
{\cal L}_{dil}=
\sqrt{-g}[(16\pi G)^{-1} {\cal R}-2(\nabla{\Phi})^2+
e^{-2\alpha\Phi}F_{\mu\nu}F^{\mu\nu}]+ {\cal L}_{matter}. 
\label{lag} 
\end{eqnarray}
Lagrangian (\ref{lag}) contains the KK and the SS theories for 
$\alpha=\sqrt{3}$ and $\alpha=1$ respectively and the Einstein-Maxwell 
one for $\alpha=0$. After a conformal transformation, the Brans-Dicke 
theory is also contained here for $\alpha=1 / \sqrt{2\omega+3}$ (where the 
parameter $\omega$ is a measure of the influence of the scalar field on the 
gravitational field). The dilaton field $\Phi$ is coupled to an 
electromagnetic potential with Faraday tensor given by $F_{\mu\nu}$. 
 
In the attempt to construct a cosmological model that keeps all the possible 
Universe components, we start from Lagrangian (\ref{lag}) and assume 
that ${\cal L}_{matter}={\cal L}_{fluid}-\sqrt{-g}[V(\Phi)+\Lambda]$, 
where ${\cal L}_{fluid}$ represents the content of ordinary matter which 
is supposed 
to be a perfect fluid with energy-momentum tensor 
$T=diag(-\rho,p,p,p)$, equation of state $p = \omega \rho$ (with $\omega$ 
constant), and will redshift as $\rho=\rho_0 a^{-3(1+\omega)}$. We have 
thus added a model for the fermionic sector to the bosonic part emerging 
from dilaton theory. We are considering the possibility of a non--vanishing 
vacuum energy density and the inclusion of $V(\Phi)$ brings the possibility 
of a phase transition in the early universe. As we said, this is feasible 
for a KK model, where there are no prescriptions for the potential at low 
energies. It has been seen that KK, as a fundamental theory, does not lead 
to successful inflation \cite{Levin}, \cite{cam}. Nevertheless, if we take 
it as a model for a higher--D theory, supplementing it with a Higgs 
potential, the differential equations of the theory reduce to the standard 
cosmological equations ((\ref{eq}) below).

In this way, Lagrangian (\ref{lag}) transforms into 
\begin{eqnarray}
{{\cal L}_{eff}} & = & \sqrt{-g}\left[
\frac{\cal R}{2}-2 (\nabla\Phi)^2-V(\Phi)
-\Lambda \right] + {\cal L}_{fluid},
\label{leff}
\end{eqnarray}
where we have taken $8\pi G = 1$ and we have dropped the electromagnetic 
interaction in order to have isotropy and homogeneity. Therefore the 
parameter $\alpha$, which gives the three different types of theories, 
is lost and they are put at the same (mathematical) level for this situation.

For a homogeneous and isotropic space we use the 
Friedmann--Robertson--Walker metric            
$
ds^2=-dt^2+a^2\left[{dr^2\over1-kr^2}+r^2(d\theta^2+\sin\theta d\varphi^2)
\right]. 
$
Then, the field equations derived from (\ref{leff}) are
\[
\ddot\phi+3H\dot\phi+{\partial V(\phi)\over\partial{\phi}}=0
\]
\begin{equation}
\frac{d\rho}{dt}+3H(\rho+p)=0
\label{eq}
\end{equation}
\[
3H^2+3{k\over a^2}=\rho+{1\over2}{\dot\phi}^2+V(\phi)+\Lambda\ ,
\]
where $H={\dot a / a}$ is the Hubble parameter, a dot means derivation 
with respect to cosmic time $t$ and we have made the following 
transformation $\Phi ={1\over 2}\phi$. 

For a potential of the form $V_{\pm}(\phi)=\lambda (\phi^2\pm\delta^2)^2$,
we were able to find exact inflationary solutions for this set of equations. 
\section{Solutions with $V_{+}(\phi)=\lambda(\phi^2+\delta^2)^2$}
The scale factor of the universe is found to be
\begin{equation}
a(t)=a_0 \sin^2 \left(\sqrt{\lambda}\,\delta\, t\right), \label{Aai'}
\end{equation}
where $\Lambda = -12\,\lambda\,{\delta}^{2}$, 
$ a_0=\sqrt {{\frac {2\,k}{\lambda\,{\delta}^{4}}}}$ and 
$\rho_0 = 12 \lambda \delta^{2} a_0$.
For this solution, we require an inflationary perfect fluid with $w=-2/3$, a 
negative cosmological constant $\Lambda$ and a closed topology of the 
universe, $i.e.$, $k=+1$. 

The scalar field is given by
\begin{equation} 
\phi (t)=  \delta \ \cot \left(\sqrt{\lambda}\,\delta\, t\right) \label{eq3}
\end{equation}

The total energy density
$\rho_{total}=\rho+{1\over2}{\dot\phi}^2+V_{+}(\phi)+
\Lambda$ (fig. 1) is dominated by the scalar field at early times, 
resulting in an inflationary period from the beginning 
to ${1\over 2}\sqrt{-\Lambda \over 3}t+{\pi\over 2}={\pi\over 4}$; at 
later times, the influence of the negative 
cosmological constant closes the universe. In between, the inflationary 
fluid may, depending on the values of the parameters of the theory, 
represent the dominant contribution.

It is interesting to notice that, even though the total energy density in the 
early universe is dominated by the scalar field, the same expansion law would 
have been obtained in a flat universe, filled just with an inflationary 
perfect fluid and the (negative) cosmological constant since, with our solution, the
full Friedmann equation can in fact be divided into two blocks in which the 
curvature term is cancelled by the scalar field energy (both potential and 
kinetic) and the Hubble parameter is equal to the perfect fluid 
and the $\Lambda$ terms. In fact, the effective cosmological constant is, 
when the scalar field has settled down, 
$\Lambda_{eff} = -12 \lambda \delta^2 + \lambda \delta^4$, which can be 
positive. Nonetheless, this does not modify the expansion since the 
positive term $(\lambda \delta^4)$ does not dictate directly the Hubble 
parameter behaviour.

The evolution of $\phi$ is such that it starts in $+\infty$, goes to its 
minimum value ($\phi = 0$) when the universe reaches the point of maximum 
expansion and then to $-\infty$. 
$\phi$ never settles down in the minimum of its potential but, for small 
values of the parameters of the theory --$\delta^2$ and the self--coupling 
constant $\lambda$, or equivalenty the value of the cosmological 
constant-- it can spend a long time near its vacuum value. This behaviour 
is shown in fig. 2.

The universe presents a cyclic behaviour with period 
$T = 4\pi \sqrt {-3\over \Lambda}$, with subsequent deflationary and 
inflationary episodes. 
$\phi$ is discontinouos at each point where $a = 0$. But, if we identify the $-\infty$ and $+\infty$ values in the 
potential $V(\phi)$, forming a ``compactified potential", then, the universe 
goes through the ``BB" passing by the identified $\pm\infty$ point (see
fig.3). Strictly speaking, the set 
$V_+=\{(x,y)\in R^2|y=(x^2+\delta^2)^2\}$ 
is non-compact, but the set $V_+^C=V_+\cup\{\infty\}$ is, if we use the 
canonical topology of compactified sets, as it is shown in fig.3. The 
motivation for this compactification comes from the graphs for
$V_{+}(\phi)$ and $\phi$, which are in the $R^2$ plane. If we compactify 
$R^2$ to $S^2$, i.e. if we bend $R^2$ untill we obtain a 2--dimensional 
sphere, the graphs for $V_{+}(\phi)$ and $\phi$ can be extended on the 
sphere, forming continuous functions. 

Notice that for the empty Lema\^{\i}tre models, i.e. non--zero 
cosmological constant, with positive curvature, only a positive $\Lambda$ is 
allowed (see e.g. \cite{OR}). Here, allowing for other components in the
universe, we were able to find an exact solution for a model with positive 
curvature and negative cosmological constant.
\section{Solutions with $V_{-}(\phi)=\lambda(\phi^2-\delta^2)^2$}
We have again solutions for an inflationary perfect fluid, with $w=-2/3$ and 
we were able to find a solution only for a universe with positive curvature. 
The universe experiences an inflationary expansion, as expected, since 
all the energy sources are inflationary.  The inflationary scalar field 
decays to $\delta$ (see below) but, due to the presence of the non--vanishing 
cosmological term, the universe never recovers from the inflationary stage. 
As in the previous case, although the influence of the perfect fluid on the 
total energy density can be felt only in a reduced epoch (see fig.4), its 
presence is necessary for the existence of our solution. The evolution of 
the Hubble parameter depends in fact, as in the previous case, from the 
$\Lambda$ and the perfect fluid terms only.  

The scale factor of the universe is now given by
\begin{equation}
a(t)= a_0 \sinh^2 \left(\sqrt{\lambda}\,\delta\, t\right) 
\label{Bai'}
\end{equation}
with $\Lambda = 12 \lambda \delta^2$, 
$a_0=\sqrt {{\frac{2\,k}{\lambda\,{\delta}^{4}}}}$ and 
$\rho_0 = 12 \lambda \delta^2 a_0$. We require $k=+1$ and a positive 
cosmological constant.

The evolution of the scalar field is 
\begin{equation}
\phi (t) = \delta \coth \left(\sqrt{\lambda}\,\delta\, t\right) .
\label{eq8}
\end{equation}
 
The scalar field starts in $+\infty$ and quickly settles to ($+\delta$), 
without passing through $\phi=0$ (see fig. 5). The potential energy comes
then from infinity to its minimum so that, although $V_{-}(\phi)$ is a
symmetry breaking potential, since $V_{-}(\phi)$ never takes the value of
its local maximum in $V_{-}(\phi=0)$, a phase transition is not achieved 
through the local maximum. For the case of an oscillating universe, the
point $t=0$ could be defined in any point where $a=0$. Here, $t=0$ 
corresponds to the only point where 
the scale factor goes to zero, defining in such a way negative times for the 
left branch, where the universe is deflationary. This phase of accelerated 
contraction preceeding the big bang, corresponds to a time reversal 
transformation from the $t > 0$ solution and can be considered as a 
particular case of the pre--big--bang scenario \cite{gasp}.
So, if we consider the negative $t$ branch, $\phi$ is settled in one of the 
true vacuum states ($-\delta$), then it climbs the potential in the 
negative $\phi$ 
value branch to $-\infty$ when $t=0$. After the BB, $\phi$ rolls down 
from $+\infty$ to $+\delta$. In such a way, the scalar field goes 
from $-\delta$ (in the pre-BB) to $+\delta$ (in our universe) through infinity 
and, if we again suppose that the $-\infty$ and $+\infty$ values are 
identified in the potential $V_{-}(\phi)$, forming a ``compactified
potential" $V_{-}^C = V_{-}(\phi) \cup \{\infty\}$, 
the universe achieves an {\it exotic} phase transition passing through
the identified $\mp \infty$ point (see fig.6).
  
Our solutions lead, in both cases, to an initial singularity, no matter how 
high is the value of $\Lambda$. Of course, this approach has limitations: in 
the vicinity of the Planck scale, the low--energy effective action does not 
apply and some modifications must be introduced. Then, following the 
suggestion in \cite{venez}, taking into account a self--dual dilaton 
potential could help avoiding the singularity, as in \cite{luca}.

Another important modification that should be introduced at very early 
epochs comes from the finite temperature correction terms.
\section{Conclusions.}
Keeping all the usually accepted possible components of the early 
universe: the curvature term, a perfect fluid, a scalar field and a 
cosmological constant term, we found two classes of exact inflationary 
solutions, with some constraints on the allowed values of the envolved 
physical quantities.
We were able to find exact solutions only for a closed universe with a 
perfect fluid with equation of state $p = -(2/3)\rho$. Then, our models are 
not realistic in the sense that they do not describe radiation or matter 
dominated epochs, nevertheless they give the possibility to follow the 
evolution of a many components universe and verify the fulfillment of the 
inflationary regime. Our solutions can be a good approximation to the
universe behaviour even at later epochs: for instance, a recent work 
from Turner and White \cite{white} shows that the best fit to all present 
cosmological data is obtained with a flat CDM universe made of 
$\Omega_{matter}\sim 0.3$, $h\sim 0.7$ and a smooth component contributing 
the remaining energy density, with equation of state $p_X=w\rho_X$, 
with $w~\sim -0.6$. Our perfect fluid with equation of state $p=-(2/3)\rho$, 
with a specific selection of the set of parameters, could have this role 
in the present universe.

The influence of the scalar field on the universe evolution basically 
dominates near the past and future singularity, then it decays and the universe 
behavior is dictated at late times by the cosmological constant. In order to 
obtain exact solutions, none of the components in our model could be dropped 
out.
In the class with negative $\Lambda$, the universe has an inflationary epoch 
with a graceful exit, and then recollapses under the influence of the 
cosmological constant. This kind of solutions does not allow a symmetry 
breaking potential. For the class of solutions with a non--vanishing positive 
cosmological term, the universe enters an inflationary stage but, as 
expected, never recovers from it. 
In both cases we have cosmologies with a big bang: even a high cosmological constant value is not able to prevent the shrink to a singularity.

Even in the case where the scalar field has a symmetry breaking potential 
$V_{-}(\phi)$, its evolution is such that it does not go through a phase 
transition;  at least in this analysis where the finite--temperature 
effects are not taken into account. 

A more complicated and interesting scenario, with several epochs of 
inflationary and deflationary regimes, arises when considering a 
pre--big--bang phase, resulting from S--duality.
Then, a possible interpretation of the evolution of the scalar field through the $t=0$ point, rendering it smooth, emerges from invoking a 
compactified potential at the transition epoch between the two dual phases. 
This compactification, in the case of the $V_-(\phi)$ potential, leads to a non--standard phase transition, that we call {\it exotic} phase tarnsition, where the scalar field goes from one minimum to the other passing through the identified $\pm \infty$ p

oint.
The initial singularity is somehow contrary to the spirit of a pre--big--bang 
cosmology, but it has to be stressed that we are using the low--energy 
effective action near the Planck scale, where in fact we are not expecting it 
to be valid, so that some modifications must be introduced in order to test 
the Universe behaviour near ``$t = 0$''.
\section{Acknowledgments}
This work is partially supported by CONACyT-M\'exico. MM is grateful to 
CONACyT for the support through fellowship Reg. No. 91825. GP acknowledges
the support from UNAM-DGAPA, IN-109896.

\newpage
\figure{
\epsfxsize = 250pt
\epsfbox{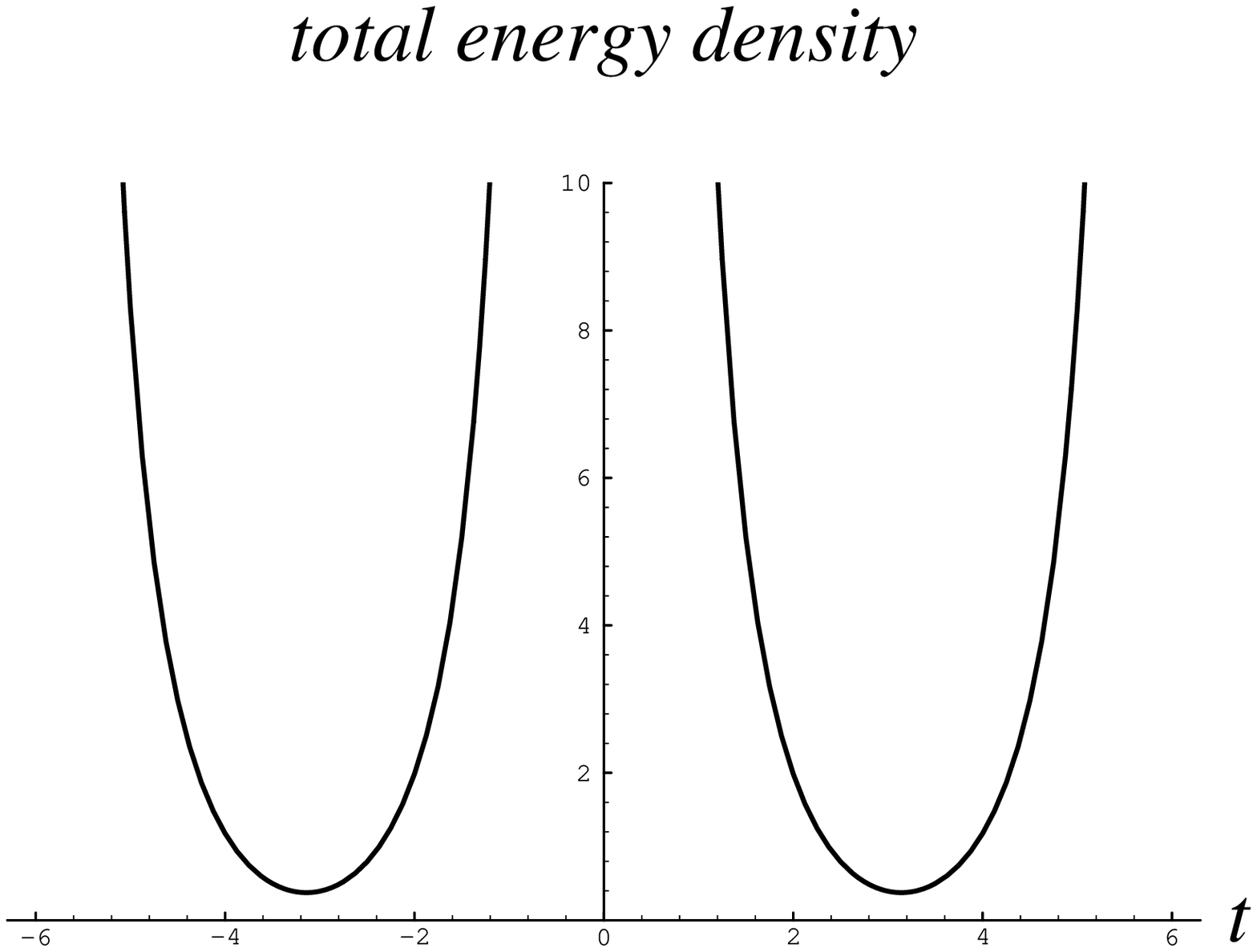}
\caption{
The total energy density $\rho_{total}= \rho +\frac12
{\dot\phi}^2 +\lambda (\phi^2+\delta^2)^2+\Lambda$ for the case
$w=-\frac23$, negative $\Lambda$ and $k=+1$, with $\lambda
=+\frac14$ and $\delta =1$.}
}
\newpage
\figure{
\epsfxsize = 250pt
\epsfbox{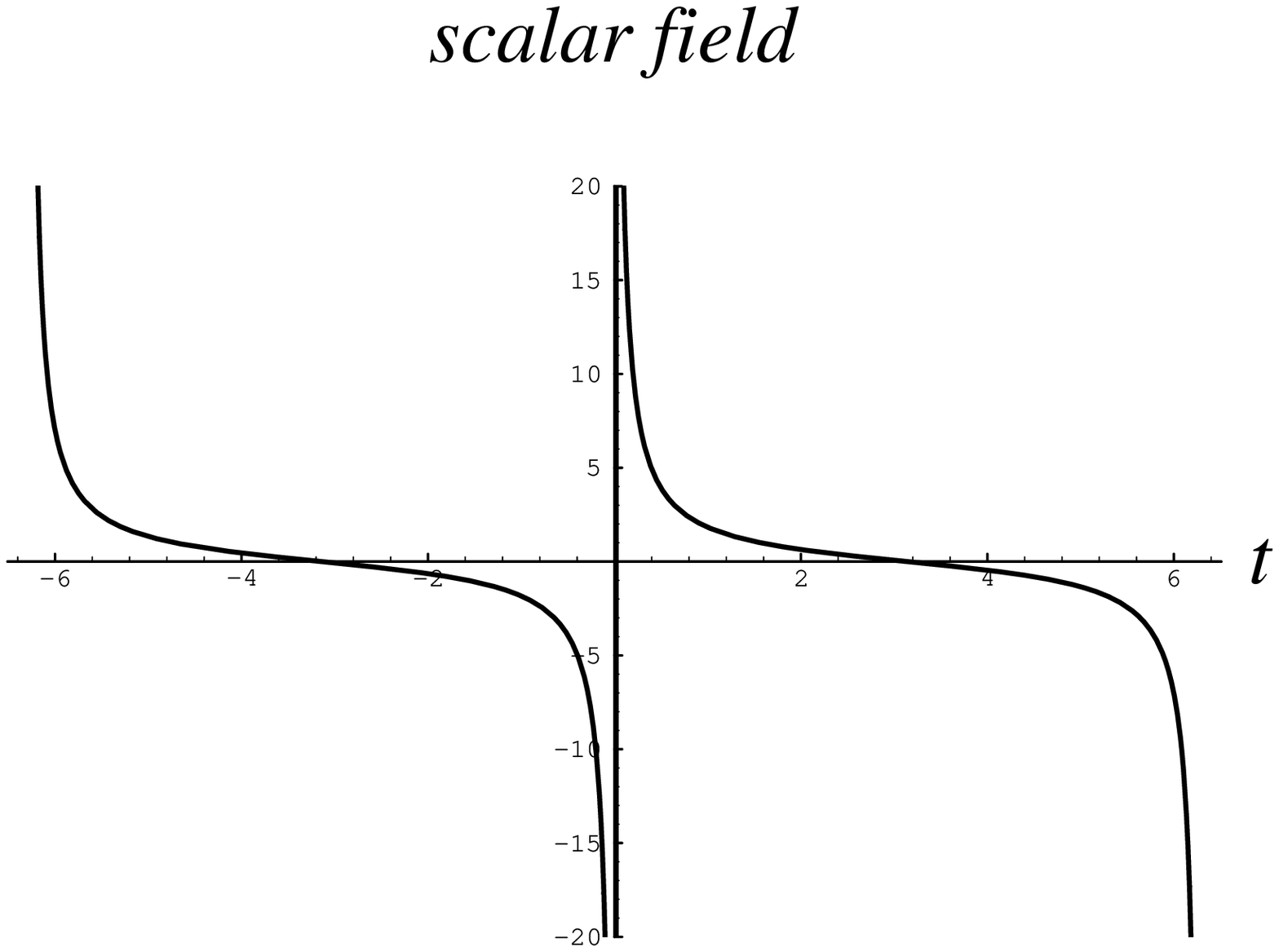}
\caption{
The scalar field $\phi$ associated to $V_{+}(\phi)$ , with $k=+1$, 
$\lambda=\frac14$ and $\delta =1$.}
}
\newpage
\figure{
\epsfxsize = 250pt
\epsfbox{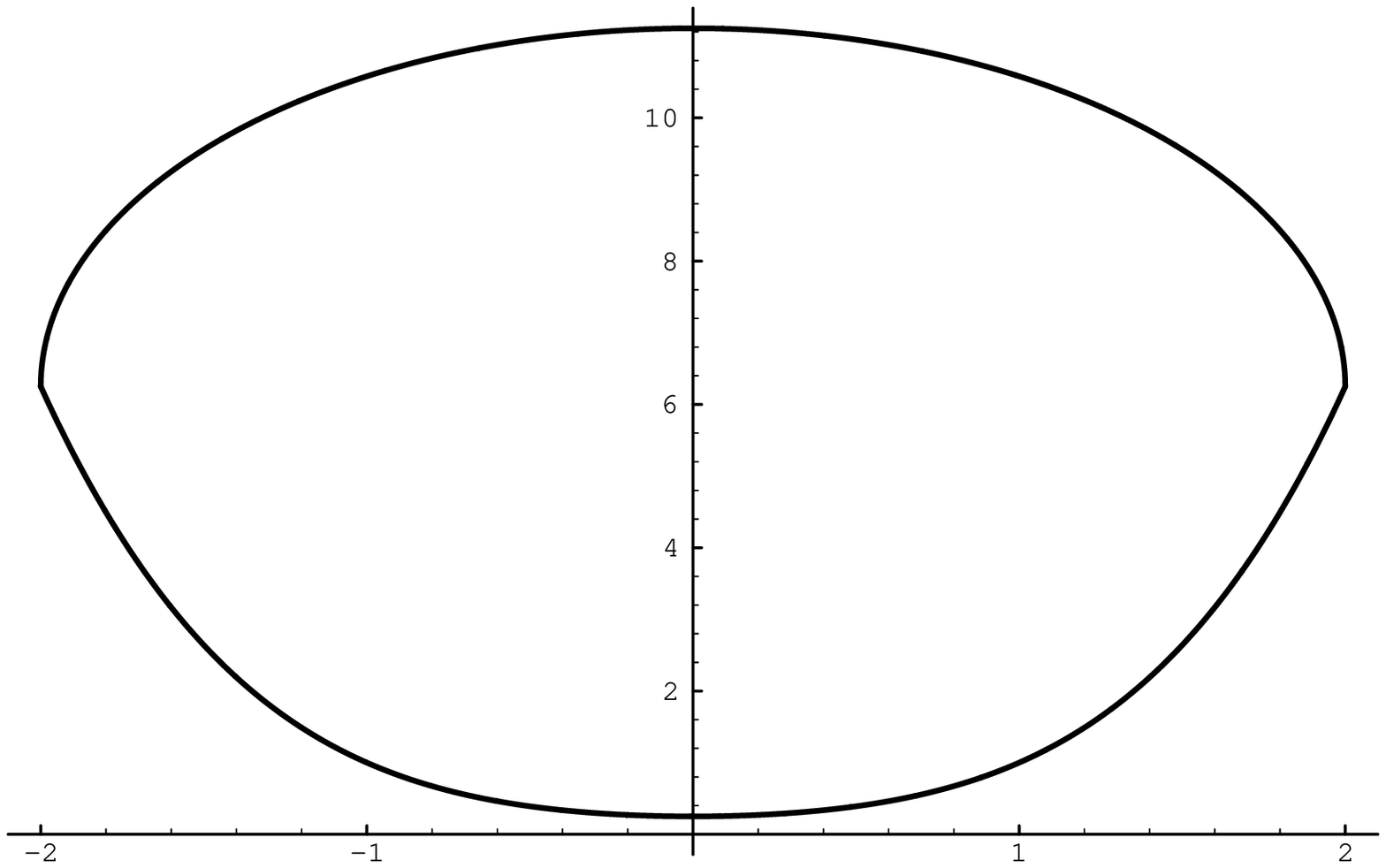}
\caption{
The compactified potential field $V_{+}^C = V_{+}(\phi)\cup
\{\infty\}=\lambda (\phi^2+\delta^2)^2 \cup \{\infty\}$, where we have
identified $-\infty$ and $+\infty$ at the $t=0$ point.}
}
\newpage
\figure{
\epsfxsize = 250pt
\epsfbox{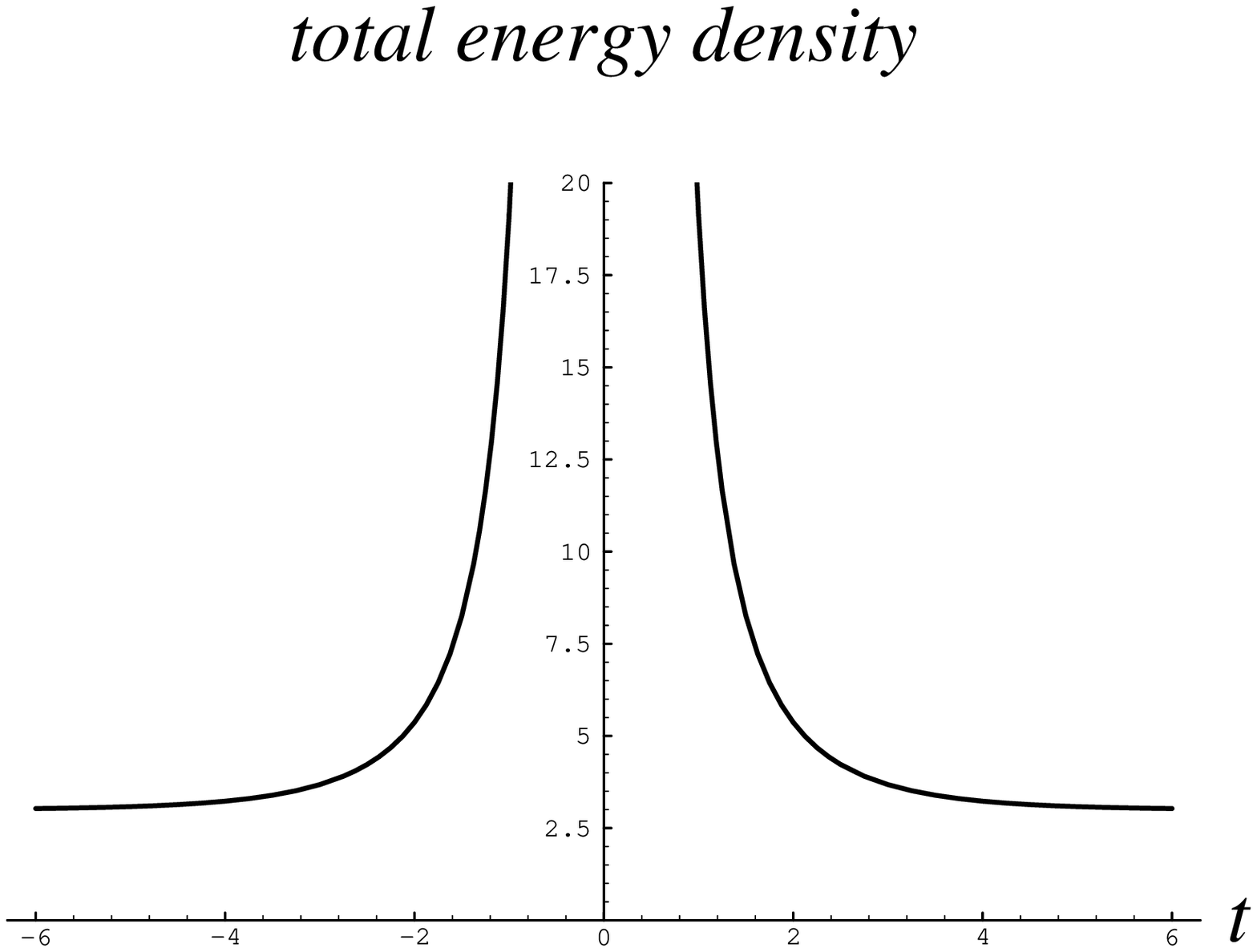}
\caption{
The total energy density $\rho_{total}=\rho+\frac12
{\dot\phi}^2+\lambda (\phi^2-\delta^2)^2+\Lambda$ for the case
$w=-\frac23$, positive $\Lambda$ and $k=+1$, with
$\lambda=\frac14$ and $\delta=1$.}
}
\newpage
\figure{
\epsfxsize = 250pt
\epsfbox{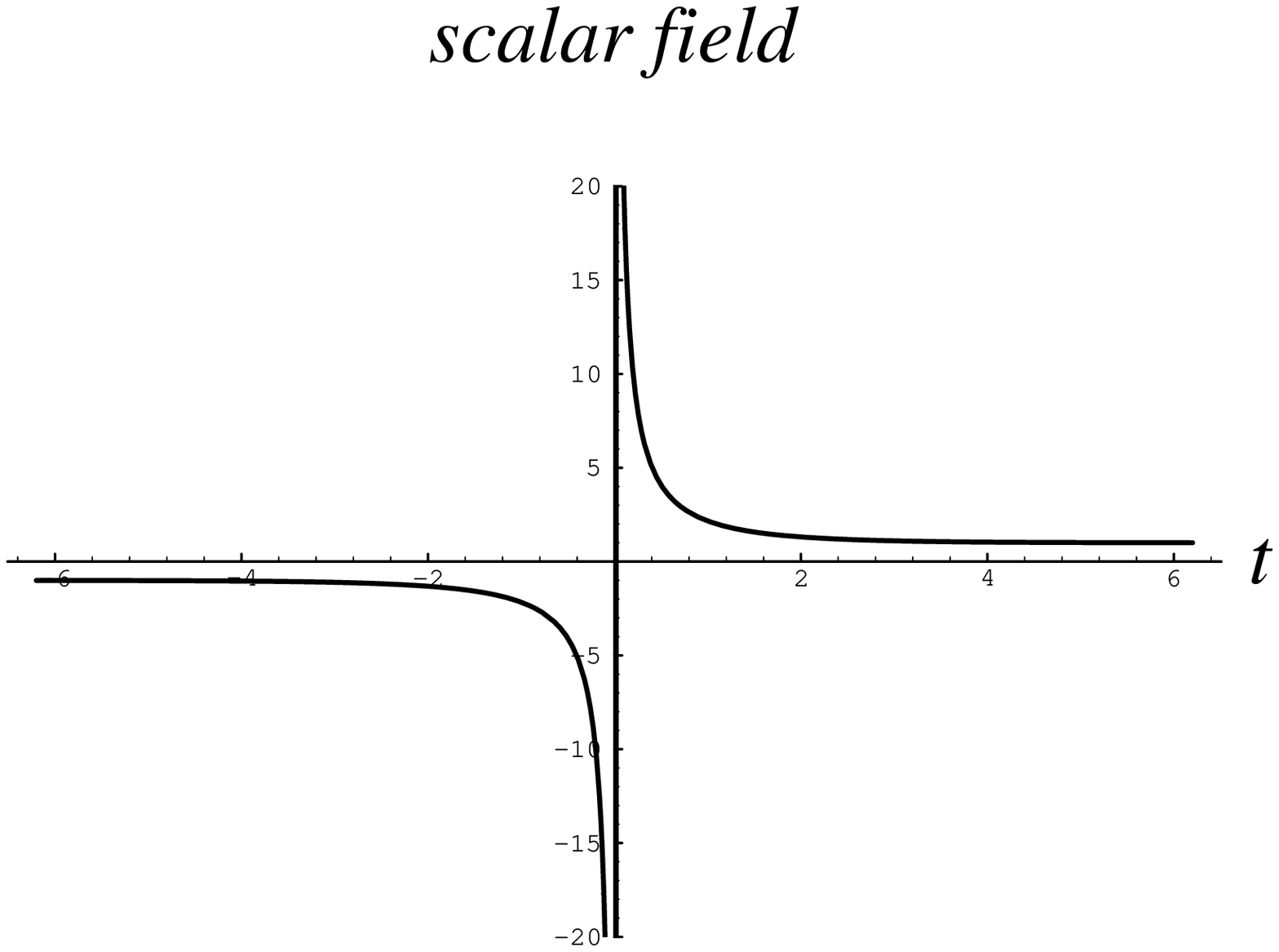}
\caption{
The scalar field $\phi$ associated to $V_{-}(\phi)$, with $k=+1$,
$\lambda =\frac14$ and $\delta =1$.}
}
\newpage
\figure{
\epsfxsize = 250pt
\epsfbox{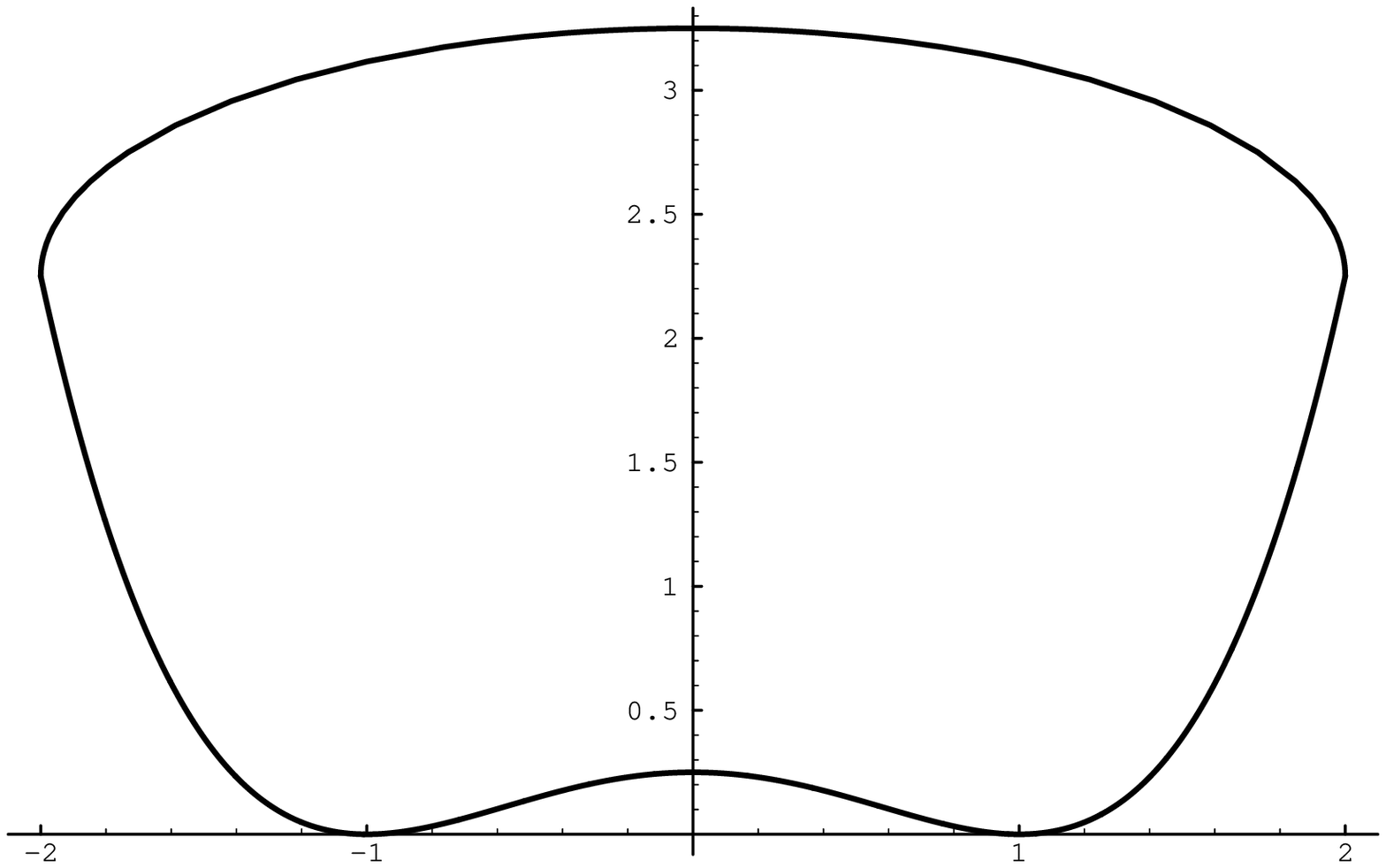}
\caption{
The compactified potential field $V_{-}^C = V_{-}(\phi) \cup
\{\infty\} = \lambda (\phi^2-\delta^2)^2 \cup \{\infty\}$, where we have
identified $-\infty$ and $+\infty$ at the $t=0$ point.}
}

\begin{thebibliography}{99}
\bibitem{linde} A. D. Linde. Particle Physics and Inflationary Cosmology. 
{\it Harwood Academic Publishers}, Chur (1990).

\bibitem{Guth} A.H. Guth. {\it Phys. Rev.} {\bf D23}, 347 (1981).

\bibitem{lucchinm} F. Lucchin and S. Matarrese, {\it Phys. Rev.} {\bf D32}, 
1316 (1985).

\bibitem{alguien} S. Kalara, N. Kaloper and K.A. Olive,  {\it Nucl. Phys.} 
{\bf B341}, 252 (1990).

\bibitem{olive} K.A. Olive, {\it Phys. Pep.} {\bf 190}, 307 (1990).

\bibitem{artlinde} A. D. Linde and A. Mezhlumian, {\it Phys.Rev.} {\bf D52}, 
6789 (1995).

\bibitem{chim} L.P. Chimento and A.S. Jakubi,    {\it Int. J. Mod. Phys.}
{\bf D5} 71 (1996). 

\bibitem{mielke} F. E. Schunck and E. W.. Mielke, {\it Phys. Rev.} 
{\bf D50}, 4794 (1994).

\bibitem{milke1} E. W. Mielke and F. E. Schunck, {\it Phys. Rev.} 
{\bf D52}, 672 (1995).

\bibitem{carroll} S.M. Carroll and W.H. Press, 
{\it Annu. Rev. Astron. Astrophys.} {\bf 30}, 499 (1992).

\bibitem{picci} B.A. Campbell, A. Linde, K.A. Olive, {\it Nucl. Phys.} 
{\bf B355}, 146 (1991).

R. Holman, E.W. Kolb, S.L. Vadas and Y. Wang, {\it Phys. Rev.} {\bf D43}, 
995 (1991).

P. Bin\'etruy and M.K. Gaillard, {\it Phys. Rev.} {\bf D34}, 3069 (1986).

\bibitem{Levin} J. Levin,  {\it Phys. Lett. } {\bf B343}, 69 (1995).   

\bibitem{broso} A. Macias, A. Camacho and T. Matos,  
{\it Int. J. Mod. Phys. } {\bf D5} 617 (1995).   

\bibitem{cho} Y M Cho, {\it J.Math.Phys.}, {\bf 16}, 2029 (1975).

\bibitem{Brustein} R. Brustein and P.J. Steinhardt, {\it Phys. Lett.} 
{\bf B302}, 196 (1993).

\bibitem{Font} A. Font, L.E. Ib\'a\~nez, D. L\"{u}st and F. Quevedo, 
{\it Phys. Lett.} {\bf B249}, 35 (1990).

\bibitem{axel} A. De la Macorra, S. Lola, {\it Phys. Lett.} {\bf B373}, 299 
(1996).

\bibitem{Cop} E.J. Copeland, A.R. Liddle, D. H. Lyth, E. D. Stewart and 
D. Wands, {\it Phys. Rev.} {\bf D49}, 6410 (1994).

\bibitem{venez} M.Gasperini and G. Veneziano, 
{\it Astropart. Phys.} {\bf 1} 317 (1993).

\bibitem{luca} C. Angelantonj, L. Amendola, M. Litterio and F. Occhionero, 
{\it Phys. Rev. D} {\bf 51} 1607 (1995). 

\bibitem{cam} A. Camacho, Master Thesis, UAM-IZtapalapa (1994).


\bibitem{OR} H. C. Ohanian and R. Ruffini, Gravitation and Spacetime, 
2nd ed. {\it W.W. Norton \& company} (1994).

\bibitem{gasp} M. Gasperini, {\it The inflationary role of the dilaton in 
string cosmology}, Proceedings of the international workshop ``Birth of the 
Universe and Fundamental Physics", Rome, 18-21 May 1994, ed. F. Occhionero,
{\it Springer--Verlag, Berlin} (1995).

\bibitem{white}
M. Turner and M. White, {\it Phys. Rev.} {\bf D56} 4439 (1997).
\end{thebibliography}
\end{document}